\documentclass[11pt]{article}
\usepackage{latexsym}
\usepackage{amssymb}
\usepackage{epsf}
\usepackage{amsmath}
\usepackage[hypertex]{hyperref}
\usepackage{graphicx}

\newcommand{\lsim}{\lesssim}
\newcommand{\gsim}{\gtrsim}
\newcommand{\tr}{{\rm Tr}}

\begin{document}
\pagestyle{empty}

\begin{flushright}
TU-927
\end{flushright}

\vspace{3cm}

\begin{center}

{\bf\LARGE Emergent Higgs from hidden dimensions} 
\\

\vspace*{1.5cm}
{\large 
Ryuichiro Kitano and Yuichiro Nakai
} \\
\vspace*{0.5cm}

{\it Department of Physics, Tohoku University, Sendai 980-8578,
 Japan}\\

\end{center}

\vspace*{1.0cm}

\begin{abstract}
{\normalsize
The Higgs mechanism well describes the electroweak symmetry breaking in
 nature. We consider a possibility that the microscopic origin of the
 Higgs field is UV physics of QCD. We construct a UV complete model of a
 higher dimensional Yang-Mills theory as a deformation of a
 deconstructed (2,0) theory in six dimensions, and couple the top and
 bottom (s)quarks to it.  We see that the Higgs fields appear as
 magnetic degrees of freedom. The model can naturally explain the masses
 of the Higgs boson and the top quark. The $\rho$ meson-like resonances
 with masses such as 1~TeV are predicted.}
\end{abstract} 

\newpage
\baselineskip=18pt
\setcounter{page}{2}
\pagestyle{plain}
\baselineskip=18pt
\pagestyle{plain}

\setcounter{footnote}{0}

\section{Introduction}

The Higgs mechanism in the Standard Model and the chiral symmetry
breaking in QCD have the same structure; the nature has broken the
electroweak symmetry twice in the same way.
This situation motivates us to consider a possibility that the UV
physics of QCD is operating as the dynamics for the electroweak symmetry
breaking.
An example of such a scenario has been proposed in
Refs.~\cite{Dobrescu:1998dg,Cheng:1999bg,ArkaniHamed:1998sj,
ArkaniHamed:2000hv} in extra dimensional theories.
If the Standard Model is defined in a compactified higher dimensional
space-time, the QCD interaction gets strongly coupled at high energy and
causes condensations of fermion pairs such as $\langle \bar t t
\rangle$, breaking the electroweak symmetry.
The low energy effective theory looks like the Standard Model where the
Higgs boson is an emergent particle.

This scenario relies on strong gauge interactions in higher dimensional
theories.
However, it is this strong coupling that requires a UV cut-off to define
a higher dimensional gauge theory.
As in the Nambu--Jona-Lasinio model \cite{Nambu:1961tp} for the chiral
symmetry breaking, whether or not a condensation forms depends crucially
on how the theory is cut-off, and thus discussion requires a UV
completion of the theory.

A UV complete definition of a higher dimensional gauge theory may, in
fact, be provided within four dimensional field theories by using the
deconstruction
technique~\cite{ArkaniHamed:2001ca,Hill:2000mu,Cheng:2001vd} and
supersymmetry (SUSY).
It has been argued in Ref.~\cite{ArkaniHamed:2001ie} that one can obtain
the (2,0) superconformal field theory in six dimensions compactified on
a torus as an $N \to \infty$ limit of a four dimensional ${\cal N}=2$
supersymmetric $SU(N_c)^N$ quiver gauge theory. The continuum limit
provides a five dimensional ${\cal N}=2$ supersymmetric gauge theory on
a circle defined with two parameters: a gauge coupling $g_5$ and a
radius $R_5$.
The Kaluza-Klein (KK) tower of gauge fields starts at $1/R_5$, and the
theory gets strongly coupled at a scale $\Lambda_5 = {8 \pi^2 /
g_5^2}$. What is interesting is that, at this scale, it starts to appear
a tower of magnetic gauge fields which can be seen from the S-duality in
four dimensional supersymmetric theories. The tower represents the KK
modes of a 6th dimension, and thus one can identify $\Lambda_5 = {1 /
R_6}$.  Thanks to the finiteness of the theory, we do not need to
cut-off the theory at this scale even though the theory gets strongly
coupled. See, for example, \cite{Seiberg:1997ax} for relations among
maximally supersymmetric theories in various dimensions, and
also~\cite{Douglas:2010iu, Lambert:2010iw} for recent approaches to the
(2,0) theory from five dimensional maximally supersymmetric theories.

The theory with a finite $N$ provides the low energy effective theory of
the (2,0) theory, that approximates the $N \to \infty$ limit up to a
scale $N/R_5$. For $1/R_6 < N/R_5$, gauge coupling constants are strong
at every site, and thus one should include magnetic degrees of freedom
in the effective theory. What is important here is that, in principle,
the deconstructed theory enables us to discuss physics beyond the scale
$\Lambda_5$ by dealing with strongly coupled four dimensional theories.

Armed with the confidence that a higher dimensional gauge theory exists
and the strong coupling simply implies the appearance of magnetic
degrees of freedom, one can now consider a possibility that the Higgs
field in the Standard Model is one of such emergent degrees of freedom.
In this paper, based on the deconstructed (2,0) theory, we propose a
two-site Standard Model as an effective theory of a higher dimensional
Standard Model where the fundamental Higgs fields are absent.
If one of the two QCD couplings is large, one can find that the Higgs
fields appear as magnetic degrees of freedom. The four dimensional QCD
interaction is kept weakly coupled since it is mainly from the other
site.
The electroweak symmetry breaking is described as the usual Higgs
mechanism as in the Minimal Supersymmetric Standard Model (MSSM).
The situation we consider is different from the models in
Refs.~\cite{Dobrescu:1998dg,Cheng:1999bg,ArkaniHamed:1998sj,
ArkaniHamed:2000hv} where it is assumed that the compactification scale
is of the order of $1/R_5 \sim$~TeV.
The scale $1/R_5$ in our framework can be much larger than TeV. The
electroweak scale is rather generated by physics at $1/R_6 (\ll 1/R_5)$
where the magnetic degrees of freedom appear.

The Standard Model in a latticized extra dimension has the structure of
the topcolor model~\cite{Hill:1991at} as noted in
Ref.~\cite{Cheng:2001nh}.
The two-site model we discuss is indeed similar to the super topcolor
model~\cite{Fukushima:2010pm}, where the emergent Higgs fields are
originally a pair of scalar top quarks.
(See also \cite{Craig:2011ev} and \cite{Csaki:2011xn} for proposals of
the Higgs fields as magnetic degrees of freedom.)
We will see that the Higgs boson and the top/bottom quarks obtain 
correct masses.
Small masses compared to the naive estimates are explained by a weakly
interacting feature of the magnetic theory. In the explicit model we
discuss, the magnetic theory is a conformal field theory, and the
Higgs/top/bottom fields interact rather weakly to the dynamical
sector. This is in fact a realization of the partially composite Higgs
scenario in SUSY proposed in Ref.~\cite{Kitano:2012wv}.

\section{Deconstructed (2,0) theory and non-supersymmetric Yang-Mills
 theory}

\renewcommand{\arraystretch}{1.3}
\begin{table}[t]
\begin{center}
\begin{tabular}[t]{l|c|c|c|c|c|c}
 & $SU(N_c)_1$ & $SU(N_c)_2$ & $SU(N_c)_3$ & $\cdots$ & $SU(N_c)_{N-1}$
& $SU(N_c)_N$ \\
\hline
$\mathcal{Q}_1$ & $N_c$ & $\overline{ N_c }$ & 1 & $\cdots$ & 1 &1 \\
$\mathcal{Q}_2$ & 1 & $N_c$ & $\overline{ N_c }$ & $\cdots$ & 1 &1 \\
$\vdots$ &  &  & & &  & \\
$\mathcal{Q}_{N-1}$ & 1 & 1 & 1 & $\cdots$ & $N_c$ & $\overline{N_c}$ \\
$\mathcal{Q}_{N}$ & $\overline{N_c}$ & 1 & 1 & $\cdots$ & 1 & $N_c$ \\
\vspace{0.0cm}
\end{tabular}
\caption{The deconstructed (2,0) theory. \label{tab:(2,0)}} 
\end{center}
\end{table}
\renewcommand{\arraystretch}{1}

\begin{figure}
\begin{center}
\includegraphics[width=5cm]{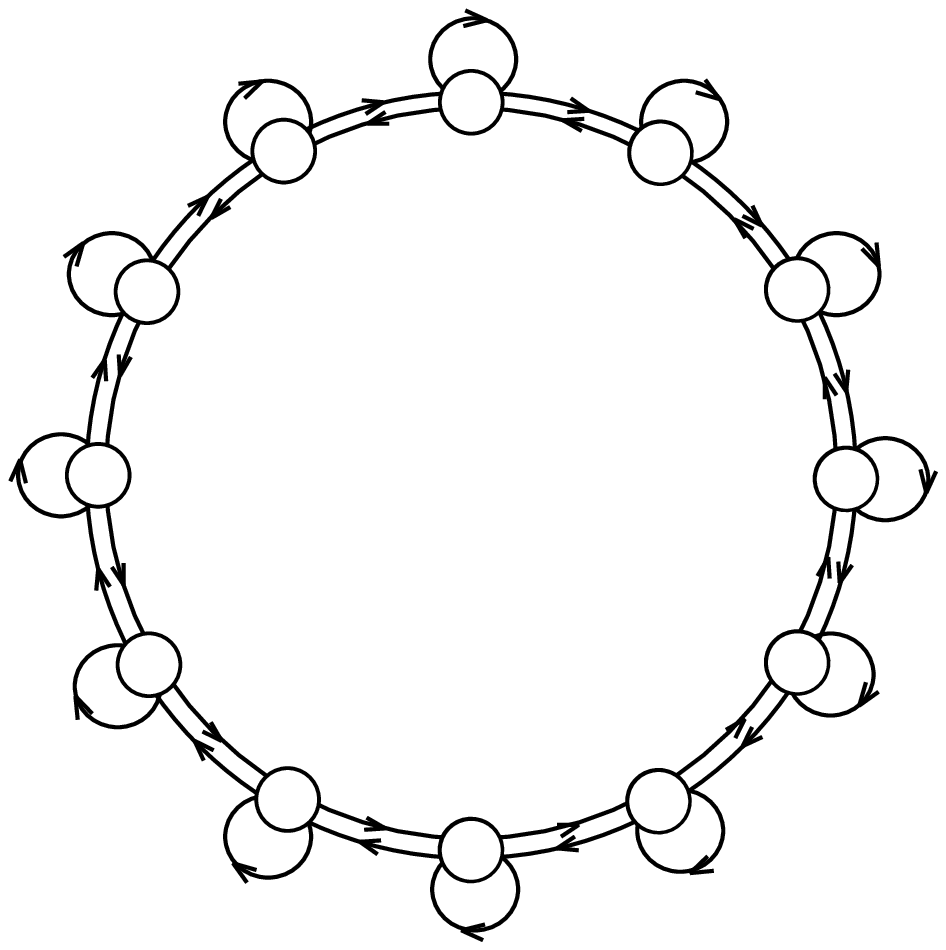}\hspace{2cm}
\includegraphics[width=5cm]{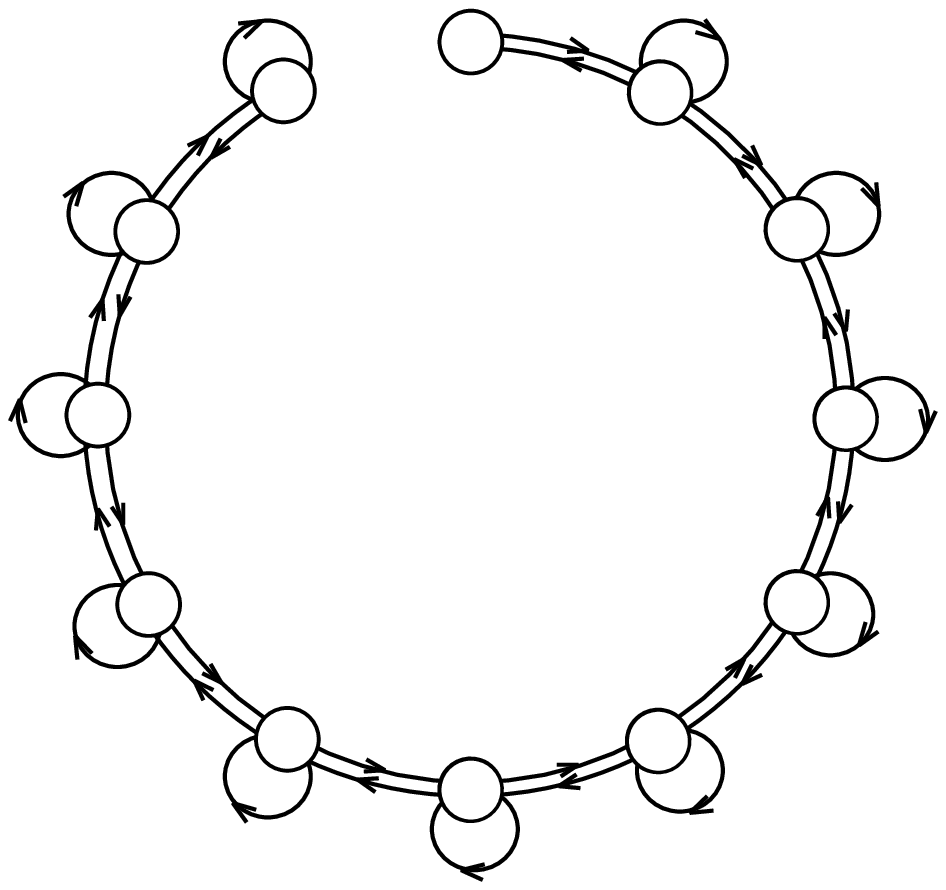} \caption{The quiver diagram
of the deconstructed (2,0) theory (left) and an ${\cal N}=1$ deformation
 (right).
\label{fig:(2,0)}}
\end{center}
\end{figure}

It has been argued that a four dimensional ${\cal N} = 2$ supersymmetric
$SU(N_c)^N$ gauge theory defines the (2,0) theory in six dimensions on a
torus in the $N \to \infty$ limit~\cite{ArkaniHamed:2001ie}.
The matter content is given in Table.~\ref{tab:(2,0)}, where
$\mathcal{Q} (= Q, \bar{Q})$'s are hypermultiplets which acquire vacuum
expectation values (VEVs), $Q = \bar Q = v {\bf 1}$. The quiver diagram
is shown in Fig.~\ref{fig:(2,0)}.
The 5th dimension can be seen as the tower of the gauge
fields, and the 6th one appears as the tower of the magnetic states
through the S-duality. The correspondence is
\begin{eqnarray}
 {1 \over R_5} = {4 \pi g v \over N},\ \ \ 
 {1 \over R_6} = { (4 \pi)^2 v \over g} = { 8 \pi^2 \over g_5^2},
\label{eq:radi}
\end{eqnarray}
where $g$ is the gauge coupling constant of each site, and
$g_5$ is the gauge coupling constant of the 5D theory.
With fixed $N$, the quiver theory is weakly coupled when $1/R_6 > N /
R_5$.

In order to take the $N \to \infty$ limit while $R_5$ and $R_6$ fixed, one
needs to send $g \to \infty$.  The claim of
Ref.~\cite{ArkaniHamed:2001ie} is that one can take this limit since the
theory is finite. One does not need to worry about the perturbativity or
a possible phase transition.

One can make an IR deformation of the model to reduce to an ${\cal N} =
1$ SUSY Yang-Mills theory. A simple way is to add masses to
$\mathcal{Q}_N$ and an adjoint chiral superfield in the ${\cal N} = 2$
gauge multiplet at the 1st site, $\Phi_1$, such as
\begin{eqnarray}
\Delta W = m Q_N \bar{Q}_N + \frac{m'}{2} \Phi_1^2.
\label{eq:deform}
\end{eqnarray}
By further adding a gaugino mass, one can obtain a
non-supersymmetric Yang-Mills theory. See Fig.~\ref{fig:(2,0)} (right)
for the quiver diagram of the ${\cal N}=1$ model.

The VEV $v$ (and thus the volume) is not fixed at this level.
One can fix this by gauging a baryon symmetry and add a superpotential
term at the 1st site,
\begin{eqnarray}
 \sqrt 2 g_B Q_1 \Phi_B \bar{Q}_1 - \sqrt 2 g_B v^2 \Phi_B,
\end{eqnarray}
where $\Phi_B$ is the singlet chiral superfield in the ${\cal N}=2$
gauge multiplet. By taking $g_B \to \infty$, the gauge multiplet becomes
a Lagrange multiplier.

In general, the deconstruction technique provides a UV complete 4D
theory which has the same IR physics as the extra dimensional model if
the coupling constant is small. The $N$-site model provides the same
physics as the 5D theory up to a scale $N/R_5$.
A lesson from the (2,0) theory is that the argument holds for strong
coupling. 
One can keep the cut-off scale as $N/R_5$ whereas the strong coupling
scale is set by another parameter $8 \pi^2 / g_5^2 = 1/R_6$ which can be
lower than $N/R_5$, i.e., we can discuss physics much above the energy
scale $8\pi^2 / g_5^2$ by including the magnetic degrees of freedom in
the effective theory.
Below, we couple matter fields in the Standard Model to the ${\cal N}=1$
Yang-Mills model, and analyze it in a strongly coupled regime.
The model is supposed to describe the IR physics of a 5D Standard Model.

\section{The two-site Standard Model}

\renewcommand{\arraystretch}{1.3}
\begin{table}[t]
\begin{center}
\begin{tabular}[t]{c|c|c|c||c}
 & $SU(3)_1$ & $SU(3)_2$ & $U(1)_B$ & $SU(2)_L$ $\times$ $U(1)_Y$ \\
\hline
$Q$ & $3$ & $\overline{ 3 }$ &${1}$ & $1_0$\\
$\bar Q$ & $\overline{ 3 }$ & $3$ & $-1$ &  $1_0$\\
$\Phi$ & 1 & $1 + 8$ & 0 &  $1_0$\\
\hline 
$q_1$ & $3$ & $1$ & $1$ & ${2}_{1/6}$\\
$t^c_1$ & $\overline{3}$ & $1$ & $-1$ & $1_{-2/3}$\\
$b^c_1$ & $\overline{3}$ & $1$ & $-1$ & $1_{1/3}$\\
\hline 
$q_2$ & 1 & $3$ & 0 & ${2}_{1/6}$\\
$t_2^c$ & 1 & $\overline{3}$& 0 & $1_{-2/3}$\\
$b_2^{c}$ & 1 & $\overline{3}$ & 0 & $1_{1/3}$\\
\hline
$\bar q_2$ & 1 & $\overline{3}$ & 0 & ${\bar 2}_{-1/6}$\\
$\bar t_2^{c}$ & 1 & ${3}$ & 0 & $1_{2/3}$\\
$\bar b_2^{c}$ & 1 & ${3}$ & 0 & $1_{-1/3}$\\
\end{tabular}
\caption{The two-site Standard Model. \label{tab:SM}}
\end{center}
\end{table}
\renewcommand{\arraystretch}{1}

\begin{figure}
\begin{center}
\includegraphics[width=5cm]{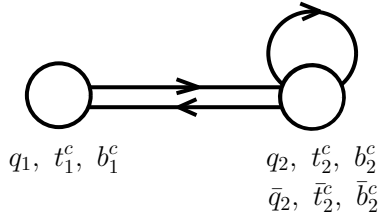} 
\end{center}
\caption{The quiver diagram of the two-site Standard
 Model. \label{fig:SM}}
\end{figure}

We now introduce matter fields in the Standard Model in the
framework. We are not aware of how we obtain such fields in a full UV
finite theory. We simply introduce them and couple to the model in the
previous section.
Since we are only interested in the low energy physics, we consider a
model with $N=2$ which can describe physics below $1/R_5$.

As in Table.~\ref{tab:SM}, we consider a model where only $SU(3)$ gauge
factor has the structure of an extra dimensional theory. (See
Fig.~\ref{fig:SM} for the quiver diagram.) This means that we have
already integrated out the KK modes of the $SU(2)_L \times U(1)_Y$ gauge
fields. We are implicitly assuming that there are no significant effects
at low energy from the $SU(2)_L \times U(1)_Y$ dynamics in the extra
dimension.
We introduce the top and bottom quark multiplets at both the 1st and the
2nd sites, representing that the matter fields are propagating in the
extra dimension~\cite{Skiba:2002nx}. In order to reproduce the correct
number of the chiral matter, vector-like partners are introduced at the
2nd site. For simplicity, we put other matter fields such as quarks in
the 1st and 2nd generations and leptons at the 2nd site for now.
Interestingly, the model is very similar to the (super) topcolor model
\cite{Hill:1991at,Fukushima:2010pm}.
The superpotential terms relevant for the discussion are
\begin{eqnarray}
 W = \sqrt 2 g \left(
q_1 \bar Q \bar q_2
+ t^c_1 Q \bar t^c_2
+ b^c_1 Q \bar b^c_2
+ \bar Q \Phi Q
- v^2 \tr \Phi
+ v_q \bar q_2 q_2
+ v_t \bar t^c_2 t^c_2
+ v_b \bar b^c_2 b^c_2
\right).
\end{eqnarray}
The mass terms at the 2nd site represent the profiles of the matter
fields in the extra dimension. For 
\begin{eqnarray}
 v_q = v_t = v_b = v,
\end{eqnarray}
the zero modes of the matter fields have flat distributions.

The gauge coupling constants at the 1st and the 2nd sites run
differently once we deform the theory. In particular, the 1st site is
asymptotically free whereas the 2nd one is IR free.
We define here the dynamical scale of $SU(3)_1$ as $\Lambda$. For 
\begin{eqnarray}
 \Lambda \ll 4 \pi v,
\end{eqnarray}
the classical level analysis is reliable. In this case, the low energy
theory is the MSSM without the Higgs fields.
The low energy QCD coupling is given by
\begin{eqnarray}
 {1 \over g_{\rm QCD}^2 } = {1 \over g_1^2} + {1 \over g_2^2},
\end{eqnarray}
where $g_1$ and $g_2$ are coupling constants of $SU(3)_1$ and
$SU(3)_2$, respectively.

Below we analyze the strongly coupled region $\Lambda \gg 4 \pi v$. We
assume that $g_2$ is small enough to reproduce the QCD coupling at low
energy.
The difference of the coupling between $g_1$ and $g_2$ may need some
explanation in order for the model to be interpreted as an IR theory of
a higher dimensional theory.
A possibility is just a quantum effect due to the different
renormalization group running. One can also assume that the gauge
coupling is position dependent. For example, there can be a large
localized kinetic term for the gauge boson somewhere away from the 1st
site.

It is not completely clear if the model with matter fields is really a
part of some UV complete higher dimensional theory, although we suspect
that the string theory construction is possible.
In any case, the two-site reduced model itself is a renormalizable
theory and its IR behavior can be reliably analyzed as we see below. One
can, of course, take the two-site model as the definition without
referring to extra dimensions.

\section{IR physics -- the Seiberg dual picture}

\renewcommand{\arraystretch}{1.3}
\begin{table}[t]
\begin{center}
\begin{tabular}[t]{c|c|c|c||c}
 & $SU(2)_1$ & $SU(3)_2$ & $U(1)_B$ & $SU(2)_L$ $\times$ $U(1)_Y$ \\
\hline
$f$ & $2$ & $1$ & $3/2$& $2_0$\\
$\bar f_u$ & $\overline{ 2 }$ & $1$ & $-3/2$&  $1_{1/2}$\\
$\bar f_d$ & $\overline{ 2 }$ & $1$ & $-3/2$&  $1_{-1/2}$\\
$H_u$ & $1$ & $1$ & 0 &  $2_{1/2}$\\
$H_d$ & $1$ & $1$ & 0 &  $2_{-1/2}$\\
\hline
$f^\prime$ & $2$ & ${ 3 }$ & $3/2$ & $1_{1/6}$\\
$\bar f^\prime$ & $\overline{ 2 }$ & $\overline{3}$ & $-3/2$&  $1_{-1/6}$\\
\hline 
$q$ & 1 & $3$ &$0$ & ${2}_{1/6}$\\
$t^c$ & 1 & $\overline{3}$ & $0$ & $1_{-2/3}$\\
$b^{c}$ & 1 & $\overline{3}$ & $0$ & $1_{1/3}$\\
\end{tabular}
\caption{The dual picture of the two-site Standard Model below the scale
$\Lambda$. \label{tab:dualeff}}
\end{center}
\end{table}
\renewcommand{\arraystretch}{1}

The theory at the 1st site is a supersymmetric QCD with $N_c = 3$ and
$N_f = 5$. This theory is at the strongly coupled edge of the conformal
window. For a strongly coupled region,
\begin{eqnarray}
 \Lambda \gg 4 \pi v,
\end{eqnarray}
the IR physics is better described by the Seiberg dual picture
\cite{Seiberg:1994pq}.
The dual picture is an $SU(2)$ gauge theory with five flavors, and has a
more weakly coupled fixed point. The particle content of the dual theory
is given in Table.~\ref{tab:dualeff}.  The fields $f$, $\bar{f}$, $f'$,
and $\bar{f}'$ are dual quarks, and $H_u (\sim q_1 b_1^c)$ and $H_d
(\sim q_1 t^c_1)$ are meson fields in the dual theory\footnote{We use
the convention which is commonly used in the MSSM literature, where
$H_u$ ($H_d$) has an hypercharge $1/2$ ($-1/2$). This is in fact a
slightly confusing notation. In this model, $H_u$ and $H_d$ are made of
the down and up-type quark superfields, respectively, and indeed, as we
will see later, $H_u$ and $H_d$ respectively give masses to the down and
up-type quarks unlike the MSSM.}. The top and bottom quarks $q$, $t^c$,
and $b^c$ are mixtures of the mesons and those at the 2nd site.
The superpotential is
\begin{eqnarray}
 W =
h \left(
  \bar f_u H_d f
+ \bar f_d H_u f
- {\epsilon}_t \bar f_u t^c f^\prime
- {\epsilon}_b \bar f_d b^c f^\prime
- {\epsilon}_q \bar f^\prime q f
\right)
+ {(4 \pi v)^2 \over \Lambda} \bar f^\prime f^\prime,
\label{eq:belowL}
\end{eqnarray}
where
\begin{eqnarray}
 {\epsilon}_{q,t,b} = {(4 \pi)^2 v_{q,t,b} \over 
\sqrt{(h \Lambda)^2 + ((4 \pi)^2 v_{q,t,b})^2 }},
\label{eq:eps}
\end{eqnarray}
and $h$ is a dimensionless coupling constant.
The scale $\Lambda$ represents the dynamical scale of the $SU(3)_1$ at
which massive hadronic modes appear. 
In general, the parameter $\Lambda$ appearing in the superpotential can
be an arbitrary value, and a choice would change the dynamical scale of
the dual theory. We take the parameter as $\Lambda$ such that the dual
theory has the same dynamical scale.  We put a factor of $(4 \pi)^2$
from the naive dimensional analysis (NDA)~\cite{Manohar:1983md, NDA}.

Since we assume $\Lambda \gg 4 \pi v$, the mass term of $f'$ and
$\bar f'$ is much smaller than $\Lambda$. Below the dynamical scale
$\Lambda$, the theory is, therefore, approximately conformally
invariant. The dimensions of the chiral superfields can be determined by
the $a$-maximization~\cite{Intriligator:2003jj}. By choosing the
parameters $v_q$, $v_t$ and $v_b$ so that $\epsilon_q \sim \epsilon_t
\sim 1$ and $\epsilon_b \ll 1$, which will be later required from $m_t
\gg m_b$, we obtain the dimensions as follows:
\begin{eqnarray}
 D(H_d) = 1.03, \ \ \ D(H_u) = 1.13,\ \ \ D(q) = 1.13,\ \ \ D(t^c) =
  1.17,
\end{eqnarray}
\begin{eqnarray}
 D(f) = 0.99, \ \ \ D(\bar f_u) = 0.99,\ \ \ D(\bar f_d) = 0.88,\ \ \ D(f') =
  0.84,\ \ \ D(\bar f^\prime) = 0.88.
\end{eqnarray}
We can see that the dimensions are close to unity, representing that the
theory is weakly coupled, and thus the effective description in terms of
the magnetic degrees of freedom is appropriate.
By matching these dimensions to the one-loop level computations of the
anomalous dimensions, we obtain the sizes of the gauge coupling and the
superpotential couplings as follows:
\begin{eqnarray}
\frac{\tilde{g}}{4 \pi} \sim 0.41,\ \ \ 
\frac{\lambda_{d}}{4 \pi} \sim 0.11,\ \ \ 
\frac{\lambda_{u}}{4 \pi} \sim 0.26,\ \ \ 
\frac{\lambda_{t}}{4 \pi} \sim 0.29,\ \ \ 
\frac{\lambda_{q}}{4 \pi} \sim 0.26,
\label{eq:lams}
\end{eqnarray}
where $\tilde g$ is the gauge coupling of $SU(2)_1$ and $\lambda$'s are
coupling constants in the superpotential:
\begin{eqnarray}
 W =
\lambda_{d}  \bar f_u H_d f
+ \lambda_u \bar f_d H_u f
+ \lambda_t \bar f_u t^c f^\prime
+ \lambda_b \bar f_d b^c f^\prime
+ \lambda_q \bar f^\prime q f
+ {(4 \pi v)^2 \over \Lambda} \bar f^\prime f^\prime,
\label{eq:fix}
\end{eqnarray}
where $\lambda_b \ll 4 \pi$. 
Since the gauge coupling $\tilde g$ is not very small, the one-loop estimation
should be regarded as an order estimate.
We see that $\lambda_d$ is somewhat smaller than others.
We keep track of these couplings $\lambda$.

The $f^\prime$ and $\bar f^\prime$ fields have a mass term with a
coefficient $(4 \pi v)^2 / \Lambda$. Through the anomalous dimension of
the $\bar f^\prime f^\prime$ operator, the actual mass scale
$\Lambda^\prime$ is slightly larger by a relation:
\begin{eqnarray}
 {(4 \pi v)^2 \over \Lambda} \left(
\Lambda \over \Lambda^\prime
\right)^{2-D(\bar f^\prime)-D(f^\prime)} = \Lambda^\prime.
\label{eq:lampdef}
\end{eqnarray}
By integrating them out, we obtain a superpotential:
\begin{eqnarray}
 W =
\lambda_d
  \bar f_u H_d f
+ \lambda_u \bar f_d H_u f
- {\lambda_q \lambda_t \over \Lambda^\prime} f \bar f_u t^c q 
- {\lambda_q \lambda_b \over \Lambda^\prime} f \bar f_d b^c q
.
\label{eq:wbelowLp}
\end{eqnarray}
At the mass scale $\Lambda^\prime$, the $SU(2)_1$ factor confines due to
the decoupling of the flavors.  The superpotential couplings $\lambda$
are now renormalized and we expect that they become larger than the
fixed point values. In the following we treat $\lambda$'s as free
parameters. 

\renewcommand{\arraystretch}{1.3}
\begin{table}[t]
\begin{center}
 \begin{tabular}[t]{c|c|c||c}
 & $SU(3)_2$ & $U(1)_B$ & $SU(2)_L$ $\times$ $U(1)_Y$ \\
\hline
$H_u$ & $1$ & 0 & $2_{1/2}$\\
$H_d$ & $1$ & 0 &  $2_{-1/2}$\\
$H^\prime_u$ & $1$ & 0 &  $2_{1/2}$\\
$H^\prime_d$ & $1$ & 0 &  $2_{-1/2}$\\
 $S$ & $1$ & 3 & 1 \\ 
 $\bar S$ & $1$ & $-3$ & 1 \\ 
\hline 
$q$ & $3$ & $0$& ${2}_{1/6}$\\
$t^c$ & $\overline{3}$ & $0$ & $1_{-2/3}$\\
$b^{c}$ & $\overline{3}$ & $0$ & $1_{1/3}$\\
\end{tabular}
\caption{The dual picture of the two-site Standard Model below
the scale $\Lambda^\prime$. \label{tab:belowLp}}
\end{center}
\end{table}
\renewcommand{\arraystretch}{1}

The low energy effective theory is described by the particles in
Table.~\ref{tab:belowLp} with a constraint:
\begin{eqnarray}
 H_u^\prime H_d^\prime - S \bar S = {\Lambda^{\prime 2} \over (4 \pi)^2}.
\label{eq:constraint}
\end{eqnarray}
A factor of $(4 \pi)^2$ is from the NDA.
The superpotential is given by
\begin{eqnarray}
 W =  {\lambda_u \Lambda^\prime \over 4 \pi} 
H_u H_d^\prime
+ {\lambda_d \Lambda^\prime \over 4 \pi} H_d H_u^\prime
- {\lambda_q \lambda_t \over 4 \pi}  H_u^\prime t^c q
- {\lambda_q \lambda_b \over 4 \pi}  H_d^\prime b^c q.
\label{eq:MSSMsupo}
\end{eqnarray}
We arrive at a model similar to the MSSM.

At the supersymmetric level, there will be no electroweak symmetry
breaking. The baryon fields $S$ and $\bar S$ develop VEVs by the
constraint in Eq.~\eqref{eq:constraint}. This will break the $U(1)_B$
and the phase direction of $S$ and $\bar S$ gets eaten by the gauge
field.
If the $U(1)_B$ factor is dynamical at this scale, there should be a
string associated with this symmetry breaking.  Since the $U(1)_B$
factor is not originally dynamical, the string should be unstable. It
implies the presence of monopoles (or dyons) at the scale $\Lambda$,
which can attach to the string. For $\Lambda \gg \Lambda'$, the string
is meta-stable.
See \cite{Kitano:2012zz} for a recent discussion on this string.

The massless degrees of freedom are just the MSSM without the Higgs
fields. That is the same result as the case with small $g$, representing
a smooth transition between strong ($\Lambda \gg 4 \pi v$) and weak
($\Lambda \ll 4 \pi v$) coupling regimes.
There are various massive modes (including a string) at the scale
$\Lambda'$. In particular, there is a light Higgs field $H_d$ with a mass
$\lambda_d \Lambda' / (4 \pi)$. This is the emergent Higgs field
responsible for the electroweak symmetry breaking as we see later.

For $\Lambda \gg 4 \pi v$, the mass scale $\Lambda^\prime$ is much
lower than $\Lambda$.
If the model has a geometric interpretation, two physical scales
$\Lambda$ and $\Lambda^\prime$ can naturally be identified as
\begin{eqnarray}
 {1 \over R_5} \sim \Lambda,\ \ \ {1 \over R_6} \sim \Lambda^\prime.
\label{eq:ident}
\end{eqnarray}
If we ignore the renormalization effects, Eqs.~\eqref{eq:lampdef} and
\eqref{eq:ident} imply $v^2 R_5 R_6$ be a constant which is indeed
the case in Eq.~\eqref{eq:radi} for a fixed $N$.

For $N_c \gg 1$, one can treat the matter fields $q$, $t^c$ and $b^c$ as
a probe since the dynamics is dominated by the $SU(N_c)$ gauge
interaction. Also, as we take the deformation in Eq.~\eqref{eq:deform}
small, the Seiberg duality is suspected to smoothly connect to the
S-duality in the ${\cal N}=2$ SUSY gauge theory~\cite{Argyres:1996eh}.
Therefore, we argue that the analysis we have done traces the properties
of the original (2,0) theory.

\section{Electroweak symmetry breaking}
As in the MSSM, adding soft SUSY breaking terms can create a vacuum at
$H_u, H_d \neq 0$, which represents the dynamical electroweak symmetry
breaking.
With generic values of the Higgs VEVs, $S$ and $\bar
S$ acquire VEVs and get eaten.

The SUSY breaking terms can be introduced in various ways. For example,
one can explicitly break SUSY at the scale $\Lambda'$ so that the
effective theory below $\Lambda'$ is just the non-supersymmetric
Standard Model. If the scale $\Lambda'$ is to be identified as the
$1/R_6$, one can consider a possibility of breaking SUSY by boundary
conditions of the 6th dimension as in Ref.~\cite{Barbieri:2000vh}.
As a similar possibility, one can introduce an $F$-component VEV of $v$
and $\Lambda$ which corresponds to the radion $F$-term, representing the
SUSY breaking by the boundary condition of the 5th
dimension~\cite{Marti:2001iw,Kaplan:2001cg}.
Through the relation in Eq.~\eqref{eq:lampdef}, $\Lambda'$ generically
acquires the $F$-component, and thus it is upgraded to a chiral
superfield $\Lambda' ( 1 + m_{\rm SUSY} \theta^2)$ where $m_{\rm SUSY}$
is a typical size of the SUSY breaking.
With the non-zero $F$-component of $\Lambda'$, we obtain a structure of
gauge mediation~\cite{Dine:1981za, Dine:1993yw, Kitano:2010fa} in
Eq.~\eqref{eq:belowL} where $f^\prime$ and $\bar f^\prime$ are messenger
fields.
For the concreteness of the discussion, we take this option of SUSY
breaking in the following.

Hereafter, we assume $m_{\rm SUSY} \sim \Lambda'$ which makes the
discussion simple and, moreover, provides a successful electroweak
symmetry breaking. We reserve a detailed analysis of the potential for a
future work.
In the following, we study the potential by treating $\lambda / 4 \pi$
as small parameters. All the estimates are based on the NDA where we
ignore all the $O(1)$ factors including color factors. The results
should be interpreted as order estimates and deviations by a factor of a
few can easily happen, although all the predictions are pretty
successful as we see below.

In the $H_u$ and $H_d$ directions, there are SUSY breaking potential
such as
\begin{eqnarray}
 V \ni 
 {m_{\rm SUSY}^2 \over (4\pi)^2}  (|\lambda_u H_u|^2 + |\lambda_d H_d|^2)
+ {1 \over (4 \pi)^2}
(|\lambda_u H_u|^4 + |\lambda_d H_d|^4).
\label{eq:pote}
\end{eqnarray}
Here we have omitted unknown $O(1)$ parameters. 
We see that the potential is suppressed by $\lambda$'s. Therefore, it is
mainly the direction of the 126~GeV Higgs boson~\cite{:2012gk, :2012gu}.
This is a realization of the partially composite Higgs scenario in
Ref.~\cite{Kitano:2012wv}.
The $H_u^\prime$ and $H_d^\prime$ direction also has a SUSY breaking
potential:
\begin{eqnarray}
 V \ni m_{\rm SUSY}^{2} (|H_u^{\prime}|^2 + |H_d^{\prime}|^2) + \cdots
\label{eq:Hprimepotential}
\end{eqnarray}
We assume that the quadratic term is positive and the soft terms are
common for $H_u'$ and $H_d'$ as they are not distinguished by the
$SU(2)_1$ gauge dynamics.

There are also contributions from the superpotential:
\begin{eqnarray}
 W \ni { \Lambda' \over 4 \pi} (\lambda_u H_u H_d' + \lambda_d H_d H_u') 
+ m_{\rm SUSY} H_u' H_d'.
\label{eq:MSSMsupo2}
\end{eqnarray}
The last term needs some explanation. It originates from the kinetic
terms of $S$ and $\bar S$, which can be expressed in terms of $H_u'$ and
$H_d'$ by using the constraint in Eq.~\eqref{eq:constraint} as
\begin{eqnarray}
 K \ni 2 \left|
H_u' H_d' - {\Lambda' \over (4 \pi)^2}
\right|.
\end{eqnarray}
By expanding around $H_u' = H_d' = 0$, we obtain
\begin{eqnarray}
K \ni \frac{\Lambda'^\dagger}{\Lambda^\prime} H_u' H_d' + {\rm h.c.}
\end{eqnarray}
The SUSY breaking term associated with this provides $\mu$- and $B
\mu$-like terms
\begin{eqnarray}
W \ni m_{\rm SUSY} H_u' H_d',\ \ \ 
 V \ni m_{\rm SUSY}^2 H_u^\prime H_d^\prime + {\rm h.c.}
\end{eqnarray}
These terms explicitly break anomalous $U(1)$ symmetries and thus
provide masses to pseudoscalar Higgs bosons.

There are also mixing terms between $H$ and $H'$ fields through $B
\mu$-like terms associated with first two terms in
Eq.~\eqref{eq:MSSMsupo2}:
\begin{eqnarray}
 V \ni m_{\rm SUSY} \left(
{\lambda_u \Lambda' \over 4 \pi} H_u H_d^\prime
+ {\lambda_d \Lambda' \over 4 \pi} H_d H_u^\prime + {\rm h.c.}
\right),
\end{eqnarray}
with which the size of the mixing is of order $\lambda / 4 \pi$.
Through these mixing terms, the $H'$ fields obtain VEVs, such as
\begin{eqnarray}
 \langle H_u' \rangle 
\sim \langle H_d' \rangle 
\sim {\lambda_d \over 4 \pi} \langle H_d \rangle
+ {\lambda_u \over 4 \pi} \langle H_u \rangle.
\label{eq:vevprime}
\end{eqnarray}

By integrating out the heavy $H'$ fields, we obtain the effective
potential of $H_u$ and $H_d$ as in the MSSM.
In the potential $V_{\rm eff} (H_u, H_d)$, at the leading order of
$\lambda/(4\pi)$, the $H_u$ and $H_d$ fields are always accompanied with
the coupling constants $\lambda_u$ and $\lambda_d$, except for the
$D$-terms of the MSSM gauge interactions.
Moreover, there is an approximate symmetry to flip $\lambda_u H_u$ and
$\lambda_d H_d$ again except for the $D$-terms and also from the quantum
corrections from the stop loops.
In this case, the potential can be written in terms of $H_+ \equiv
(\lambda_d H_d + \lambda_u H_u)/ \sqrt{2}$ and $H_- \equiv (\lambda_d
H_d - \lambda_u H_u)/ \sqrt{2}$, and that is approximately
invariant under $H_\pm \to \pm H_\pm$.
Therefore, the minimum can generically be at $\langle H_+ \rangle \neq 0$
and $\langle H_- \rangle \sim 0$, which we assume to be the case motivated
by the discussion of the $T$-parameter later.
The correction from the stop loop is of order
\begin{eqnarray}
 \langle H_- \rangle \sim 
{\lambda_q^2 \lambda_t^2 \over (4 \pi)^4} {m_{\tilde t}^2 \over m_{\rm
SUSY}^2} \langle H_+
  \rangle,
\label{eq:h-stop}
\end{eqnarray}
where $m_{\tilde t}$ is the stop mass, which is of order
\begin{eqnarray}
 m_{\tilde t}^2 \sim {\lambda_t^2 \over (4 \pi)^2} m_{\rm SUSY}^2.
\end{eqnarray}
The correction from the $D$-term is
\begin{eqnarray}
 \langle H_- \rangle \sim 
{(4 \pi)^2 \over \lambda_d^2}{m_Z^2 \over m_{\rm SUSY}^2} \langle H_+
  \rangle,
\label{eq:h-D}
\end{eqnarray}
for $\lambda_u \gg \lambda_d$. 
Higher order corrections are of order
\begin{eqnarray}
 \langle H_- \rangle \sim {\lambda_u^2 \over (4 \pi)^2} \langle H_+
  \rangle,
\label{eq:h-hi}
\end{eqnarray}
for $\lambda_u \gg \lambda_d$.
As we will see later, the VEV $\langle H_- \rangle$ should be smaller
than $\langle H_+ \rangle$ by a factor of a few. We therefore assume
that is the case, and concentrate on the $H_+$ direction, i.e., $\tan
\beta = \langle H_u \rangle / \langle H_d \rangle = \lambda_d /
\lambda_u$. For $\lambda_u \gg \lambda_d$, motivated by
Eq.~\eqref{eq:lams}, the VEV direction is approximately $H_d$, which we
assume in the following discussion.

In order to achieve the correct electroweak symmetry breaking, we need
to have the correct size of the quadratic term:
\begin{eqnarray}
{\lambda_d^2 \over (4 \pi)^2} m_{\rm SUSY}^2
\sim {m_h^2 \over 2},
\label{eq:quad}
\end{eqnarray}
where $m_h$ is the observed Higgs boson mass, 126~GeV. The
left-hand-side is a collection of various contributions such as from the
superpotential, the K{\" a}hler potential and also from a mixing with
$H'$.
The quartic term should satisfy:
\begin{eqnarray}
 {\lambda_d^4 \over (4\pi)^2}
+ {g_L^2 + g_Y^2 \over 2}
 \sim
  {m_h^2 \over \langle H \rangle^2} \sim 0.5,
\label{eq:quart}
\end{eqnarray}
in order to obtain the correct Higgs VEV, $\langle H \rangle = 174$~GeV.
The second term is from the $D$-term potential of the MSSM gauge
interactions, where $g_L$ and $g_Y$ are coupling constants of the
$SU(2)_L$ and $U(1)_Y$ gauge interactions.
Eq.~\eqref{eq:quart} implies
\begin{eqnarray}
 {\lambda_d \over 4 \pi} \sim 0.2.
\end{eqnarray}
The size is not far from the fixed point value in Eq.~\eqref{eq:lams}.
The Higgs VEV should be limited by
$
 \lambda_d \langle H_d \rangle \lesssim \Lambda^\prime,
$
since otherwise the confinement scale should be redefined as
$\Lambda^\prime \sim \lambda_d \langle H_d \rangle$. 
Therefore,
\begin{eqnarray}
 \Lambda' \gsim 400~{\rm GeV} \cdot \left(
\lambda_d / 4 \pi \over 0.2
\right).
\end{eqnarray}

The relation in Eq.~\eqref{eq:quad} now provides a measure for a
required degree of fine-tuning in the electroweak symmetry breaking such
as
\begin{eqnarray}
 \delta = 
{m_h^2 /2 \over 
(\lambda_d m_{\rm SUSY} / 4 \pi)^2,
}
=
20\% \cdot
\left(
m_{\rm SUSY} \over 1~{\rm TeV}
\right)^{-2}
\left(
\lambda_d / 4 \pi \over 0.2
\right)^{-2}
.
\end{eqnarray}
There is essentially no fine-tuning for $m_{\rm SUSY} \lesssim$~TeV.
We see that a natural electroweak symmetry breaking is achieved while
$m_h = 126$~GeV is obtained.
From the superpotential in Eq.~\eqref{eq:MSSMsupo2}, the Higgsino masses
are generated. The lightest Higgsino obtains a mass of order
\begin{eqnarray}
 m_{\tilde h} \sim
{\lambda_u \lambda_d \over (4 \pi)^2} {\Lambda'^2 \over m_{\rm SUSY}}
\sim 120~{\rm GeV} \cdot 
\left(
\lambda_d / 4 \pi \over 0.2
\right)
\left(
\lambda_u / 4 \pi \over 0.6
\right)
\left(
\Lambda^\prime \over 1~{\rm TeV}
\right)^2
\left(
m_{\rm SUSY} \over 1~{\rm TeV}
\right)^{-1}
.
\end{eqnarray}
In the case where the gauginos are much heavier than the Higgsinos, the
charged Higgsino and the neutral Higgsino degenerate. The Higgsino mass
bounds in such a case are $m_{\tilde h} \gtrsim 90$~GeV, obtained from
the searches for mono-photon signals at the LEP-II experiments
\cite{Heister:2002mn,Abdallah:2003xe}.
Although this is an order estimate, one can expect a quite light
Higgsino if there is no significant fine-tuning in the electroweak
symmetry breaking.

The top quark mass is generated by two sources. One is from the Yukawa
interaction in the superpotential in Eq.~\eqref{eq:MSSMsupo} together
with the VEV of $H_u'$ in Eq.~\eqref{eq:vevprime}. There is also a
contribution from a K{\" a}hler term:
\begin{eqnarray}
 K \ni {\lambda_q \lambda_t \lambda_d \over (4 \pi)^2} 
{1 \over \Lambda'^{\dagger}}
 H_d^\dagger t^c q,
\end{eqnarray}
which generates the non-holomorphic Yukawa term through SUSY breaking.
Both contributions are of the order of
\begin{eqnarray}
 m_t \sim {\lambda_q \lambda_t \lambda_d \over (4 \pi)^2}  \langle H_d
  \rangle
\sim 160~{\rm GeV} \cdot
\left(
\lambda_d / 4 \pi \over 0.2
\right)
\left(
\lambda_q / 4 \pi \over 0.6
\right)
\left(
\lambda_t / 4 \pi \over 0.6
\right)
.
\label{eq:mtnp}
\end{eqnarray}
The top quark mass, $m_t = 173$~GeV, can be successfully explained. The
bottom quark mass is also generated in the same way. The correct size
can be obtained by choosing the value of $\epsilon_b$ appropriately.

It is interesting that the top quark is not required to be very heavy
unlike the top condensation models~\cite{Miransky:1988xi,
Bardeen:1989ds, Hill:1991at}.
The main Higgs directions, $H_u$ and $H_d$, are originally composite of
the stop and anti-stop (or sbottom and anti-sbottom). The fermion pairs
such as the $ t \bar t$ and $b \bar b$ condensations correspond to the
$F$-component of $H_u$ and $H_d$. From the equations of motion,
$F_{H_u}^\dagger = {- \partial W / \partial H_u}$ and $F_{H_d}^\dagger =
{- \partial W / \partial H_d}$, we can see that they correspond to
$H_d^\prime$ and $H_u^\prime$. The electroweak symmetry breaking in this
model is mainly by the stop condensation, and the $t \bar t$
condensation is a minor contribution. This explains why the top quark
can be light unlike non-SUSY models.

Superparticles except for Higgsinos can be much heavier than the
electroweak scale. 
The stop and sbottom masses are 
\begin{eqnarray}
 m_{\tilde t} \sim m_{\tilde b} 
\sim {\lambda_q \over 4 \pi} m_{\rm SUSY}
\sim 600~{\rm GeV} \cdot \left(
\lambda_q / 4 \pi \over 0.6
\right) 
\left(
{m_{\rm SUSY} \over 1~{\rm TeV}}
\right).
\end{eqnarray}
For a light Higgsino, the stop (sbottom) can decay into $t \chi^0$ ($b
\chi^0$) or $b \chi^+$ ($t \chi^-$). At the LHC experiments, chargino
decays are invisible due to a small mass splitting. In this case, the
search for two $b$-jets and missing transverse momentum put the severest
constraint. The current lower bound on the sbottom mass is about 600~GeV
by assuming a 100\% branching fraction into $b
\chi^0$~\cite{ref:atlas-sbottom, ref:cms-sbottom}. Our prediction is
close to this bound.
Unlike the models based on the SUSY desert, we do not expect a large
logarithmic enhancement in the quantum corrections of the SUSY breaking
terms.
In that case, the gauginos and other squarks (including the right-handed
sbottom) can be as heavy as 2~TeV while maintaining naturalness of the
electroweak symmetry breaking~\cite{Kitano:2006gv, Papucci:2011wy}.
Such a spectrum is obtained in a scenario where SUSY is broken somewhere
away from the 1st site.

\section{Fermion masses}

At this stage, the quarks in the first two generations and leptons are
massless. A possible way to couple the fermions to the Higgs fields is
to let them propagate into the extra dimension and write down
superpotential terms such as
\begin{eqnarray}
 {q_1 b_1^c F F_u^c \over \Lambda_F}
+  {q_1 t_1^c F F_d^c \over \Lambda_F}
\end{eqnarray}
at the 1st site, where $F$, $F_u^c$ and $F_d^c$ are quark and lepton
superfields.  $\Lambda_F$ is a mass scale where these terms are
generated.  These correspond to the Yukawa interaction terms at low
energy:
\begin{eqnarray}
 {\lambda_u \Lambda \over (4 \pi)^2 \Lambda_F}
\left(
\Lambda \over \Lambda'
\right)^{1-D(H_u)}
{H_u F F_u^c}
+  {\lambda_d \Lambda \over (4 \pi)^2 \Lambda_F} 
\left(
\Lambda \over \Lambda'
\right)^{1-D(H_d)}
{H_d F F_d^c}
\end{eqnarray}
By appropriately choosing the mixing factors, one can obtain a hierarchy
of the fermion masses naturally. 
Note here that the scale $\Lambda_F$ can be much higher than TeV since
$D(H_u)$ and $D(H_d)$ are close to unity.

A similar operator 
\begin{eqnarray}
  {F F_u^c F F_d^c \over \Lambda_F}
\end{eqnarray}
would cause flavor changing neutral current processes. However, if the
wave function profiles explain the Yukawa hierarchy, the flavor
violation also has the structure of the square of the Yukawa coupling
constants. This is a realization of the minimal flavor
violation~\cite{Chivukula:1987py,Hall:1990ac,D'Ambrosio:2002ex}.

\section{Phenomenology}

The properties of the Higgs boson are the same as the Standard Model
ones as long as we look at the low energy effective theory.
However, there can be contributions from physics at the scale
$\Lambda^\prime$.  Since the Higgs fields couple to charged particles
with masses of order $\Lambda^\prime \sim$~TeV, there should be imprints
of the strongly coupled sector in the property of the Higgs boson.

The charged particles with masses of order $\Lambda'$ may give an
important contribution to the $h \rightarrow \gamma \gamma$ decay.  From
the NDA analysis, the effective operator obtained by integrating out
these particles are
\begin{eqnarray}
\mathcal{L}_{\rm eff} \sim \frac{e^2}{(4\pi)^2} \frac{\lambda_d^2\langle
 H
 \rangle}{\Lambda^{'2}} \cdot h F^{\mu\nu} F_{\mu\nu}.
\end{eqnarray}
The decay width is then given by
\begin{eqnarray}
\Gamma (h \rightarrow \gamma \gamma) \propto \left| A_{\rm SM}  
+ \frac{\lambda_d^2\langle H \rangle^2}{\Lambda^{'2}}  \right|^2,
\end{eqnarray}
where $A_{\rm SM} \sim -6.5$ is the contribution from the Standard
Model.
The deviation from the Standard Model prediction, $\Delta \Gamma /
\Gamma_{\rm SM}$, is therefore estimated to be
\begin{eqnarray}
 | \Delta \Gamma  / \Gamma_{\rm SM} |
\sim {3\%}
\cdot
\left(
\lambda_d / 4 \pi \over 0.2
\right)^2
\left(
\Lambda' \over 1~{\rm TeV}
\right)^{-2}.
\end{eqnarray}
The sign is not determined. Depending on $O(1)$ parameters, one may be
able to see this contribution.

The $S$, $T$ parameters~\cite{Peskin:1990zt} provide constraints on
contributions from the $\Lambda'$ scale physics. The NDA estimate of the
$S$ and $T$ parameters are:
\begin{eqnarray}
 \Delta S \sim {1 \over \pi} \left(
\lambda_d \langle H \rangle \over \Lambda'
\right)^2
\sim 0.06 \cdot \left(
\lambda_d / 4 \pi \over 0.2
\right)^2
\left(
\Lambda' \over 1~{\rm TeV}
\right)^{-2},
\end{eqnarray}
\begin{eqnarray}
 \Delta T &\sim& {1 \over 16\pi s_W^2 c_W^2 m_Z^2}
{(\lambda_d \langle H \rangle)^4 \over \Lambda'^2}
\left(
\langle H_- \rangle \over \langle H_+ \rangle
\right)^2
\nonumber \\
&\sim&
0.09 \cdot
\left(
\lambda_d / 4 \pi \over 0.2
\right)^4
\left(
\Lambda' \over 1~{\rm TeV}
\right)^{-2}
\left(
{\langle H_- \rangle / \langle H_+ \rangle \over 0.4}
\right)^2
.
\end{eqnarray}
The experimental bounds are $\Delta S, \Delta T \lsim 0.1$.
The $S$ parameter constraint is not very strong.
The $T$ parameter is proportional to $\langle H_- \rangle^2$ which
represents the violation of the custodial symmetry. By using the
estimates in Eqs.~\eqref{eq:h-stop}, \eqref{eq:h-D}, and
\eqref{eq:h-hi}, one can see that the constraint from the $T$ parameter
is satisfied without any fine-tuning.

At the scale $\Lambda^\prime$, there are $\rho$ meson-like vector
resonances which have the same quantum numbers as the $W$ and $Z$
bosons.
The mixing with the $W$ and $Z$ bosons are expected to be of order, $g_L
/ (4 \pi)$, where $g_L$ is the gauge coupling constant of the $SU(2)_L$
gauge boson.
Through this mixing, the resonance can be produced at the LHC via $q
\bar q \to W^\prime$ or $Z^\prime$ process.
There are various decay modes. For example for the $W'$ boson, it can
decay into $t \bar b$, $\chi^0 \chi^+$, and a pair of a neutral
boson ($h$, $Z$, $H^0$) and a charged boson ($W$, $H^+$). The branching
ratios are weighted by numbers of degrees of freedom and
$\lambda^4$. For example,
\begin{eqnarray}
 {B(W^\prime \to WZ) 
\over 
B(W^\prime \to t \bar b)} \sim {\lambda_d^4 \over
 2 \lambda_q^4 N_c}.
\end{eqnarray}
The most promising decay mode to look for the resonance is $W^\prime \to
WZ$ followed by decays into three leptons. 
The experimental bound on such a resonance has been studied in
Ref.~\cite{Bellazzini:2012tv}. 
Due to a small mixing and a small branching fraction, there seems to be
no experimental constraint on this resonance for $\Lambda' \sim $~TeV
yet, but there will be a chance to see it in near future.

\section{Discussion}

The analysis of the two-site model supports that the dynamical
electroweak symmetry breaking by QCD is possible in an extra dimensional
theory. Not only that, we find that the model is phenomenologically
attractive. One can explain the 126~GeV Higgs boson mass without a
fine-tuning.
These successes give a good motivation to construct a model in string or
M-theory. Such a construction will make it completely clear if the
emergent Higgs scenario can be embedded to a UV complete higher
dimensional theory.

If we extend the model to an $N$-site model, there should appear a tower
of Higgs sectors and techni-rho mesons by sequentially taking the
Seiberg dualities. The model then looks like the 5D MSSM where $SU(2)$
gauge fields and the Higgs sector are propagating into an extra
dimension.
This suggests an interesting duality; a 5D Higgsless MSSM with bulk QCD
is dual to another 5D MSSM with the bulk electroweak sector
including the Higgs fields.

The two-site model we studied has the same structure as the one in
Ref.~\cite{Kitano:2011zk}, where it is tempted to understand the $\rho$
meson in QCD as the magnetic gauge boson.
This exhibits a similarity of the chiral symmetry breaking in QCD and
the electroweak symmetry breaking.
Our construction of the extra dimensional model provides a unification
of these two phenomena.

SUSY is probably essential for defining a higher dimensional gauge
theory. Not only for a theory definition, SUSY explains why the
Higgs boson is so light.
However, in our discussion, a light Higgs boson is obtained by a small
coupling constant rather than a low SUSY breaking scale.
From the bottom up point of view, the point is the presence of a UV
completion in which the Higgs operators do not have a potential at tree
level, and couple to some TeV dynamics with the strength of $\lambda / 4
\pi \sim 0.2$ rather than unity.
We are not aware of, but there can be such examples of
non-supersymmetric models.

Whether or not the Higgs fields are fundamental is probably not a
physical question.
Once we have a theory, they are just effective degrees of freedom
wherever they come from.
An important message from this study is that a trial to define a higher
dimensional theory leads us to a picture of emergent Higgs fields.
A generic prediction of the framework is then a presence of relatively
light resonances at a scale $\Lambda^\prime \sim 1~{\rm TeV}$. If the
Higgs boson is weakly coupled as suggested by its mass, the techni-rho
mesons should be lighter than the naive estimate, $\Lambda' < 4 \pi
\langle H \rangle \sim 2-3$~TeV. We expect to see them at the LHC.

\section*{Acknowledgments}

We would like to thank N.~Yokoi for discussions.  
RK is supported in part by
the Grant-in-Aid for Scientific Research 23740165 of JSPS.  
YN is supported by JSPS Fellowships for Young
Scientists.


\begin{thebibliography}{1}

\bibitem{Dobrescu:1998dg} 
  B.~A.~Dobrescu,
  Phys.\ Lett.\ B {\bf 461}, 99 (1999)  [hep-ph/9812349].  
  
\bibitem{Cheng:1999bg} 
  H.~-C.~Cheng, B.~A.~Dobrescu and C.~T.~Hill,
  Nucl.\ Phys.\ B {\bf 589}, 249 (2000)  [hep-ph/9912343].  
  
\bibitem{ArkaniHamed:1998sj} 
  N.~Arkani-Hamed and S.~Dimopoulos,
  Phys.\ Rev.\ D {\bf 65}, 052003 (2002)  [hep-ph/9811353].  

\bibitem{ArkaniHamed:2000hv} 
  N.~Arkani-Hamed, H.~-C.~Cheng, B.~A.~Dobrescu and L.~J.~Hall,
  Phys.\ Rev.\ D {\bf 62}, 096006 (2000)  [hep-ph/0006238].
  
\bibitem{Nambu:1961tp} 
  Y.~Nambu and G.~Jona-Lasinio,
  Phys.\ Rev.\  {\bf 122}, 345 (1961).
  
\bibitem{ArkaniHamed:2001ca} 
  N.~Arkani-Hamed, A.~G.~Cohen and H.~Georgi,
  Phys.\ Rev.\ Lett.\  {\bf 86}, 4757 (2001)  [hep-th/0104005].

\bibitem{Hill:2000mu} 
  C.~T.~Hill, S.~Pokorski and J.~Wang,
  Phys.\ Rev.\ D {\bf 64}, 105005 (2001)  [hep-th/0104035].

\bibitem{Cheng:2001vd} 
  H.~-C.~Cheng, C.~T.~Hill, S.~Pokorski and J.~Wang,
  Phys.\ Rev.\ D {\bf 64}, 065007 (2001)  [hep-th/0104179].

\bibitem{ArkaniHamed:2001ie} 
  N.~Arkani-Hamed, A.~G.~Cohen, D.~B.~Kaplan, A.~Karch and L.~Motl,
  JHEP {\bf 0301}, 083 (2003)
  [hep-th/0110146].
  
\bibitem{Seiberg:1997ax} 
  N.~Seiberg,
  Nucl.\ Phys.\ Proc.\ Suppl.\  {\bf 67}, 158 (1998)
  [hep-th/9705117].

\bibitem{Douglas:2010iu} 
  M.~R.~Douglas,
  JHEP {\bf 1102}, 011 (2011)
  [arXiv:1012.2880 [hep-th]].

\bibitem{Lambert:2010iw} 
  N.~Lambert, C.~Papageorgakis and M.~Schmidt-Sommerfeld,
  JHEP {\bf 1101}, 083 (2011)
  [arXiv:1012.2882 [hep-th]].

\bibitem{Hill:1991at} 
  C.~T.~Hill,
  Phys.\ Lett.\ B {\bf 266}, 419 (1991).

\bibitem{Cheng:2001nh} 
  H.~-C.~Cheng, C.~T.~Hill and J.~Wang,
  Phys.\ Rev.\ D {\bf 64}, 095003 (2001)
  [hep-ph/0105323].

\bibitem{Fukushima:2010pm} 
  H.~Fukushima, R.~Kitano and M.~Yamaguchi,
  JHEP {\bf 1101}, 111 (2011)  [arXiv:1012.5394 [hep-ph]].
  
\bibitem{Craig:2011ev} 
  N.~Craig, D.~Stolarski and J.~Thaler,
  JHEP {\bf 1111}, 145 (2011)
  [arXiv:1106.2164 [hep-ph]].

\bibitem{Csaki:2011xn} 
  C.~Csaki, Y.~Shirman and J.~Terning,
  Phys.\ Rev.\ D {\bf 84}, 095011 (2011)
  [arXiv:1106.3074 [hep-ph]].

\bibitem{Kitano:2012wv} 
  R.~Kitano, M.~A.~Luty and Y.~Nakai,
  JHEP {\bf 1208}, 111 (2012)
  [arXiv:1206.4053 [hep-ph]].

\bibitem{Skiba:2002nx} 
  W.~Skiba and D.~Tucker-Smith,
  Phys.\ Rev.\ D {\bf 65}, 095002 (2002)
  [hep-ph/0201056].

\bibitem{Seiberg:1994pq} 
  N.~Seiberg,
  Nucl.\ Phys.\ B {\bf 435}, 129 (1995)  [hep-th/9411149].
  
\bibitem{Manohar:1983md} 
  A.~Manohar and H.~Georgi,
  Nucl.\ Phys.\ B {\bf 234}, 189 (1984).

\bibitem{NDA}
  M.~A.~Luty,
  Phys.\ Rev.\  {\bf D57}, 1531-1538 (1998)
  [hep-ph/9706235].
  
\bibitem{Intriligator:2003jj} 
  K.~A.~Intriligator and B.~Wecht,
  Nucl.\ Phys.\ B {\bf 667}, 183 (2003)  [hep-th/0304128].

\bibitem{Kitano:2012zz} 
  R.~Kitano, M.~Nakamura and N.~Yokoi,
  Phys.\ Rev.\ D {\bf 86}, 014510 (2012)
  [arXiv:1202.3260 [hep-ph]].

\bibitem{Argyres:1996eh} 
  P.~C.~Argyres, M.~R.~Plesser and N.~Seiberg,
  Nucl.\ Phys.\ B {\bf 471}, 159 (1996)
  [hep-th/9603042].

\bibitem{Barbieri:2000vh} 
  R.~Barbieri, L.~J.~Hall and Y.~Nomura,
  Phys.\ Rev.\ D {\bf 63}, 105007 (2001)
  [hep-ph/0011311].

\bibitem{Marti:2001iw} 
  D.~Marti and A.~Pomarol,
  Phys.\ Rev.\ D {\bf 64}, 105025 (2001)  [hep-th/0106256].
  
\bibitem{Kaplan:2001cg} 
  D.~E.~Kaplan and N.~Weiner,
  hep-ph/0108001.
  
\bibitem{Dine:1981za} 
  M.~Dine, W.~Fischler and M.~Srednicki,
  Nucl.\ Phys.\ B {\bf 189}, 575 (1981);
  S.~Dimopoulos and S.~Raby,
  Nucl.\ Phys.\ B {\bf 192}, 353 (1981);
  M.~Dine and W.~Fischler,
  Phys.\ Lett.\ B {\bf 110}, 227 (1982);
  M.~Dine and W.~Fischler,
  Nucl.\ Phys.\ B {\bf 204}, 346 (1982);
  C.~R.~Nappi and B.~A.~Ovrut,
  Phys.\ Lett.\ B {\bf 113}, 175 (1982);
  L.~Alvarez-Gaume, M.~Claudson and M.~B.~Wise,
  Nucl.\ Phys.\ B {\bf 207}, 96 (1982);
  S.~Dimopoulos and S.~Raby,
  Nucl.\ Phys.\ B {\bf 219}, 479 (1983).

\bibitem{Dine:1993yw} 
  M.~Dine and A.~E.~Nelson,
  Phys.\ Rev.\ D {\bf 48}, 1277 (1993)
  [hep-ph/9303230];
  M.~Dine, A.~E.~Nelson, Y.~Nir and Y.~Shirman,
  Phys.\ Rev.\ D {\bf 53}, 2658 (1996)
  [hep-ph/9507378];
  M.~Dine, A.~E.~Nelson and Y.~Shirman,
  Phys.\ Rev.\ D {\bf 51}, 1362 (1995)
  [hep-ph/9408384].

\bibitem{Kitano:2010fa} 
For a recent review, see
  R.~Kitano, H.~Ooguri and Y.~Ookouchi,
  Ann.\ Rev.\ Nucl.\ Part.\ Sci.\  {\bf 60}, 491 (2010)
  [arXiv:1001.4535 [hep-th]].

\bibitem{:2012gk} 
  G.~Aad {\it et al.}  [ATLAS Collaboration],
  Phys.\ Lett.\ B {\bf 716}, 1 (2012)  [arXiv:1207.7214 [hep-ex]].
  
\bibitem{:2012gu} 
  S.~Chatrchyan {\it et al.}  [CMS Collaboration],
  Phys.\ Lett.\ B {\bf 716}, 30 (2012)  [arXiv:1207.7235 [hep-ex]].

\bibitem{Heister:2002mn} 
  A.~Heister {\it et al.}  [ALEPH Collaboration],
  Phys.\ Lett.\ B {\bf 533}, 223 (2002)  [hep-ex/0203020].

\bibitem{Abdallah:2003xe} 
  J.~Abdallah {\it et al.}  [DELPHI Collaboration],
  Eur.\ Phys.\ J.\ C {\bf 31}, 421 (2003)  [hep-ex/0311019].
 
\bibitem{Miransky:1988xi} 
  V.~A.~Miransky, M.~Tanabashi and K.~Yamawaki,
  Phys.\ Lett.\ B {\bf 221}, 177 (1989).

\bibitem{Bardeen:1989ds} 
  W.~A.~Bardeen, C.~T.~Hill and M.~Lindner,
  Phys.\ Rev.\ D {\bf 41}, 1647 (1990).

\bibitem{ref:atlas-sbottom}
The ATLAS Collaboration, ATLAS-CONF-2012-165.

\bibitem{ref:cms-sbottom}
The CMS Collaboration, CMS PAS SUS-12-028.
    
\bibitem{Kitano:2006gv} 
  R.~Kitano and Y.~Nomura,
  Phys.\ Rev.\ D {\bf 73}, 095004 (2006)
  [hep-ph/0602096].

\bibitem{Papucci:2011wy} 
  M.~Papucci, J.~T.~Ruderman and A.~Weiler,
  JHEP {\bf 1209}, 035 (2012)
  [arXiv:1110.6926 [hep-ph]].

\bibitem{Chivukula:1987py} 
  R.~S.~Chivukula and H.~Georgi,
  Phys.\ Lett.\ B {\bf 188}, 99 (1987).

\bibitem{Hall:1990ac} 
  L.~J.~Hall and L.~Randall,
  Phys.\ Rev.\ Lett.\  {\bf 65}, 2939 (1990).

\bibitem{D'Ambrosio:2002ex} 
  G.~D'Ambrosio, G.~F.~Giudice, G.~Isidori and A.~Strumia,
  Nucl.\ Phys.\ B {\bf 645}, 155 (2002)  [hep-ph/0207036].

\bibitem{Peskin:1990zt} 
  M.~E.~Peskin and T.~Takeuchi,
  Phys.\ Rev.\ Lett.\  {\bf 65}, 964 (1990);
  Phys.\ Rev.\ D {\bf 46}, 381 (1992).

\bibitem{Bellazzini:2012tv} 
  B.~Bellazzini, C.~Csaki, J.~Hubisz, J.~Serra and J.~Terning,
  JHEP {\bf 1211}, 003 (2012)
  [arXiv:1205.4032 [hep-ph]].

\bibitem{Kitano:2011zk} 
  R.~Kitano,
  JHEP {\bf 1111}, 124 (2011)
  [arXiv:1109.6158 [hep-th]].

\end{thebibliography}
\end{document}